\documentclass[twocolumn]{aastex63}
\shorttitle{DPNNet-2.0}
\shortauthors{Auddy et al.}
\graphicspath{{./}{figures/}}

\begin{document}

\title{DPNNet-2.0 Part I: Finding hidden planets from simulated images of protoplanetary disk gaps}

\correspondingauthor{Sayantan Auddy}
\email{sauddy@iastate.edu, sayantanauddy21@gmail.com}

\author[0000-0003-3784-8913]{Sayantan Auddy}
\affiliation{Institute of Astronomy and Astrophysics, Academia Sinica, Taipei 10617, Taiwan}
\affiliation{Department of Physics and Astronomy, Iowa State University, Ames, IA, 50010, USA}

\author[0000-0002-0786-7307]{Ramit Dey}
\affiliation{School of Mathematical and Computational Sciences, Indian Association for the Cultivation of Science, 
Kolkata-700032, India}

\author[0000-0002-8597-4386]{Min-kai Lin}
\affiliation{Institute of Astronomy and Astrophysics, Academia Sinica, Taipei 10617, Taiwan}
\affiliation{
Physics Division, National Center for Theoretical Sciences, Taipei 10617, Taiwan}

\author[0000-0002-8138-0425]{Cassandra Hall}
\affiliation{Department of Physics and Astronomy, University of Georgia, Athens, GA 30602, USA}
\affiliation{Center for Simulational Physics, University of Georgia, Athens, GA 30602, USA}



\begin{abstract} 
The observed sub-structures, like annular gaps, in dust emissions from protoplanetary disk are often interpreted as signatures of embedded planets. Fitting a model of planetary gaps to these observed features using customized simulations or empirical relations can reveal the characteristics of the hidden planets. However, customized fitting is often impractical owing to the increasing sample size and the complexity of disk-planet interaction. 
In this paper we introduce the architecture of DPNNet-2.0, second in the series after DPNNet \citep{aud20}, designed using a Convolutional Neural Network ( CNN, here specifically ResNet50) for predicting exoplanet masses directly from simulated images of protoplanetary disks hosting a single planet. 
DPNNet-2.0 additionally consists of a multi-input framework that uses both a CNN and multi-layer perceptron (a class of artificial neural network) for processing image and disk parameters simultaneously.
This enables DPNNet-2.0 to be trained using images  directly, with the added option of considering disk parameters (disk viscosities, disk temperatures, disk surface density profiles, dust abundances, and particle Stokes numbers) generated from disk-planet hydrodynamic simulations as inputs. 
This work provides the required framework and is the first step towards the use of computer vision (implementing CNN) to directly extract mass of an exoplanet from planetary gaps observed in dust-surface density maps by telescopes such as the Atacama Large (sub-)Millimeter Array. 



\end{abstract}


\section{Introduction}
Exoplanet surveys using different techniques \citep{Hoj19,deeg18,but17} have revealed a remarkable variety of planets and planetary systems throughout the galaxy \citep{cas12,bat13}. With more than 4,000 confirmed detections,  and an even larger number of detected objects awaiting confirmation, we have a rich sample of data characterizing the demographics of exoplanets. However, most of these discovered exoplanets are much older ($\sim 10^{3}$Myr). To get a complete picture of the planet formation process and its formation environment \citep[][]{ray20} it is imperative to constrain the distribution of sizes and diversity in radii, composition, and mass \citep[e.g.,][]{win15,ful17} of young planets including those at their formation stage.


%

However, with current search techniques \citep{fis14,lee18} it is particularly challenging to detect young planets embedded within their natal disk.
The spectral characteristic and the light curves from such systems are too faint compared to the emission from the young host star and the protoplanetary disk (hereafter PPDs). Thus, direct detection in the visible and infrared spectrum from $H\alpha$ \citep{cug19,zur20} or thermal emission from the gradually cooling planets embedded within the disks are rare as well as difficult. These factors add up to severely limit the detection of young exoplanets during their formation epoch. 



An alternative approach is to search for indirect influences/signatures of these unseen planets on the physical structure of the PPDs induced by disk-planet interaction. Theoretical models have long predicted disk-planet interactions \citep{goldreich80,lin93} resulting in spiral features and/or annular gaps in gas and dust density distribution \citep{lin86}. However, it is only recently that high resolution observations in sub-mm interferometry using the Atacama Large (sub-)Millimeter Array (ALMA) and other facilities (for example SPHERE, Gemini) have reported such complex - possibly planet-induced - features with unprecedented precision \citep[e.g.,][]{Clarke2018High-resolutionAu,dipierro2018,liu19,per18,Ands18,Hua18b,Huac,long18,long20,van19}. 

Near-infrared images from ALMA has revealed the ubiquity of concentric rings and gaps in many observed systems, like HL Tau, \citep{ALMA15}, TW Hya \citep{and16,Hua18a}, HD 97048 \citep{van17}, and HD 169141 \citep{mom15}. There are many possible explanations for the formation of these observed structures.
Disk-specific mechanisms such as secular gravitational instability \citep{you11,tak14}, magnetorotational instability (MRI) turbulence \citep[e.g.][]{joh09,sim14}, self-induced dust pile-ups \citep{gon15}, gap-opening in the dust spatial distribution due to large-scale vortices \citep{bar17}, radially variable magnetic disk winds \citep{sur18} are some of the possible scenarios that can lead to the formation of these substructures in a dusty disk. However, perhaps the most popular interpretation is that these substructures are induced by gap-opening planets \citep[][]{Rosotti2016TheObservations,Dipierro2016TwoDiscs,dong17}. 


Recent detection of planetary signatures `velocity kinks' using gas kinematic measurements \citep{teague2018,pinte2018,pinte19,pinte20},  particularly inside the observed dust gaps, provide strong evidence in support of planet-induced gaps. More so, direct detection of an accreting planet, using $\rm {H} \alpha$ emissions, inside the gap/cavity of the transition disk PDS-70 \citep{kep18,2018ApJ...863L...8W,2019NatAs...3..749H} strengthen the idea of planetary gaps. This further suggests that planets may form much earlier than previously thought as most of these PPDs are relatively young $<10 \rm{Myr}$. This opens up a unique window to probe unseen, young and fully grown exoplanets by characterizing substructures (particularly axisymmetric disk gaps) in PPDs from dust emission.



Disk-planet interactions are studied extensively using generic hydrodynamical simulations to reproduce dust distributions which are comparable with observations \citep{Zhang2018TheInterpretation}. Planet-induced gap features, mostly width and depth, are modeled both using numerical and analytic approaches  \citep[e.g.][]{crida06, paardekooper09, Duffell2013GAPDISK, Fung2014HOWPLANETS,Duffell2015ADISKS,Kanagawa2015MASSSTRUCTURES,ilee2020}. 
In order to infer planet properties (such as planet mass) one typically needs to match observations with outputs from these customized dusty disk-planet simulations.
Each simulation is characterized using multiple parameters, like local disk aspect ratio, disk density profile, dimensionless viscosity parameter, particle size and dust abundance (or metallicity) along with the gap features: gap depth and width. Furthermore, as the disk models improves with the inclusion of additional physics, like migrating planets \citep{par10,meru19}, multiple gaps and rings \citep{fer20}, improved thermodynamics \citep{miranda19,miranda20}, more parameters are necessary to setup the simulation. Thus, it becomes impractical to perform disk-planet simulations spanning such a large ever-increasing parameter space to match each observed target in the current and future surveys of PPDs. Thus it is essential to seek a generic, future-proof and  accurate model for planet-induced gaps.



\subsection{DPNNet-1.0}\label{DPNNet}

In \cite{aud20} (hereafter Paper 1) we introduced Disk Planet Neural Network (DPNNet), a deep learning model which predicts the planet mass from simulated/observed disk properties. It out-performs simple analytic models \citep[for example, see][]{kan16,lod19} in its ability to encapsulate the multi-dimensional parameter space, which is always a challenge with analytic relations. 
DPNNet was trained with synthetic data, normalized using the standard scaling (z-score) by removing the mean and then scaling it by the standard deviation, generated from numerical simulations using the \textsc{fargo3d}  hydrodynamics code \citep{llambay19}.

DPNNet's architecture consists of a fully connected multi-layer perceptron (MLP), which takes inputs directly from the user to predict the planet mass (see Figure 1 in Paper 1).  For disks with observed gaps, one of the key input feature is the dust gap width. The specific way in which the gap widths are measured (or the definition of gap width) varies somewhat across the literature. In general, the gap width is defined as the radial distance between the inner and the outer edge of the gap where the surface density reaches a predefined threshold, $\Sigma_{\rm T}$. In Paper 1, consistent with \cite{kan16}, we set $\Sigma_{\rm T}$ to half of the undepleted initial surface density. This differs from \cite{dong17}, where $\Sigma_{\rm T}$ is geometric mean of the undepleted and the minimum surface density, as well as \cite{Zhang2018TheInterpretation}, where $\Sigma_{\rm T}$ is the average of the peaks and the minimum (gap) surface density. Similarly, different functional forms were used by \cite{Clarke2018High-resolutionAu} (difference of two logistic functions) and \cite{long18} (symmetrical Gaussian) to fit the radial dust profile to measure the gap width. Such variations limit the application of DPNNet, requiring the network to be trained each time for different definitions of gap width. 

Additionally, DPNNet only considers the gap width, while ignoring other morphological features like gap depth \citep{Fung2014HOWPLANETS}, asymmetries like vortices \citep{val07,ham17,ham19} and spirals \citep{Zhu_2015}. Quantifying these features using user defined probes can be a complex task and further creates scope for more uncertainties. An advanced approach is to use computer vision to initiate image-based characterization of the disk features.  
This facilitates the extraction of complex non-linear features from images directly without explicit user interference, thereby minimizing the associated errors.





\subsection{Introduction to DPNNet-2.0}
In this paper we introduce DPNNet-2.0, a deep neural network model based on the Convolutional Neural Network (CNN) architecture,  to predict mass of super-Earth to Saturn-sized planets from simulated images of PPDs. 
Compared to DPNNet (with MLP architecture) which requires users to measure disk features from images, for instance obtaining the gap widths as defined by the user, DPNNet-2.0 is an end-to-end model that learns the linear and the non-linear features directly from the disk images. This  minimizes biases associated with different user-defined measurement probes. 
Additionally, DPNNet-2.0 includes a complementary module that allows it to accept disk parameters (feature variables) like disk aspect ratio, disk viscosity, disk surface density profile, metallicity, and particle size in order to better constrain the disk properties. This gives DPNNet-2.0 the flexibility to accept both images and disk attributes simultaneously. As we will demonstrate in this paper, the use of mixed data input improves the model performance while trained on the same dataset.    

\subsection{Paper plan}

The paper is planed as follows. In Section \ref{sec-simulation} we describe the disk-planet hydro-simulations as well as the parameter space considered for the current study. We introduce the CNN architecture implemented in DPNNet-2.0 along with the multi-input module in Section \ref{sec-cnn} and discuss in details the steps used to pre-process the data. In Section \ref{sec-results} we present the prediction from our trained DPNNet-2.0, based on simulated images, using both the single and the multi-input module. Finally, in section \ref{sec-discussion} we discuss the limitations and future scopes.

\begin{table*}
\centering
\caption{\textbf{Parameter space for hydrodynamic simulations.}}
\begin{tabular}{|l|l|l|l|l|l|}

 \hline
 \hline
  Name & Notation  & max.  & min.  & mean & std \\
  
 \hline

		\hline
		Planet mass in Earth masses & $M_{\rm P}/ M_\Earth$& 120 .0 & 8.0 & 64.0 & 32.3 \\
		Disk aspect-ratio &	$ h_{\rm 0}$  &  0.100  & 0.025 & 0.063 & 0.022   \\
		Disk surface density profile & $ \sigma $  &  1.50  & 0.50 & 0.75 & 0.29  \\
        Disk viscosity parameter & $\alpha$ ($\times 10^{-3}$) &  10 & 0.10  & 5.00 & 2.86 \\
		Global dust-to-gas ratio &$\epsilon$ & 0.10  & 0.01 & 0.06 & 0.03  \\
		Particle Stokes numbers& $ S_{\rm t}$  &  0.100  & 0.001 & 0.051 & 0.029 \\ 

		\hline
	\end{tabular}
\label{tab:input_data}

\end{table*}

\section{Disk-Planet simulations} \label{sec-simulation}
We are interested in modeling the structural variations of the dust surface density in a dusty protoplanetary disk due to an embedded planet. We develop a machine learning (ML) algorithm to identify and classify these variations. Our ML model is trained using sample images generated from hydrodynamic simulations of disk-planet systems. The details of the ML model and the training step are given in the following section. The physical disk model and the simulations used to generate the training data are discussed in details in Paper 1. However, for convenience of the reader we briefly review the relevant details.

\subsection{Disk Setup}
We consider a razor-thin 2D disk with a non-migrating planet of mass $M_{\rm P }$ placed at $R=R_{0}$, on a circular Keplerian orbit around a central star of mass $M_{*}$. The planet's orbital period $P_0 = 2\pi/\Omega_\mathrm{K0}$ is used as the unit of time, where $\Omega_\mathrm{K} = \sqrt{GM_*/R^3}$ is the Keplerian frequency. The sub-script $0$ denotes evaluation at the planet's location $R_{\rm 0}$. We set $G=R_0=M_*=1$, where $G$ is the gravitational constant, such that the units are all dimensionless. The disk is initialized with a gas surface density profile given as

\begin{equation}
    \Sigma_{\rm g}(R) = \Sigma_{\rm g0} \left(\frac{R}{R_{0}}\right)^{- \sigma},
\end{equation}
where $\sigma$ is the exponent of the density profile and $\Sigma_{\rm {g0}}= 10^{-4}$ in code units. The disk's self-gravity is negligible as our models are gravitationally stable with a Toomre parameters $\mathcal{Q}_{0} \gtrsim 80$. 
The disk is locally isothermal having a sound-speed profile $c_{\rm s}(R)=h_{0}R\Omega_{\rm K}$ and a constant flaring index $F$, such that aspect ratio is 

\begin{equation}
    h = \frac{H}{R} = h_{0} \left(\frac{R}{R_{0}}\right)^{F},
\end{equation}
where $H$ is the disk scale height, $h_{0}$ is the aspect ratio at the planet's location $R = R_{0}$.
The disk turbulence is modelled using the standard $\alpha-$prescription of \cite[][]{shakura73}, where the kinematic viscosity $\nu = \alpha c_{\rm s}^2/ \Omega_{\rm K}$ and $\alpha$ is a constant.

We consider gas and dust simulations where the dust particles are modeled as a pressureless fluid \citep{jacquet11}. The dust particles are parameterized using constant Stokes numbers $S_{\rm} \equiv t_{\rm s}\Omega_{\rm K}$, where the dust-gas coupling \citep[][]{weiden77} is characterized using the stopping time $t_{\rm s}$. The dust surface density is initialized as $\Sigma_{\rm d} = \epsilon \Sigma_{\rm g}$, where $\epsilon$ is the constant dust-to-gas ratio. We include the dust back reaction onto the gas.

\begin{figure*} 
\centering
\includegraphics[scale=.18]{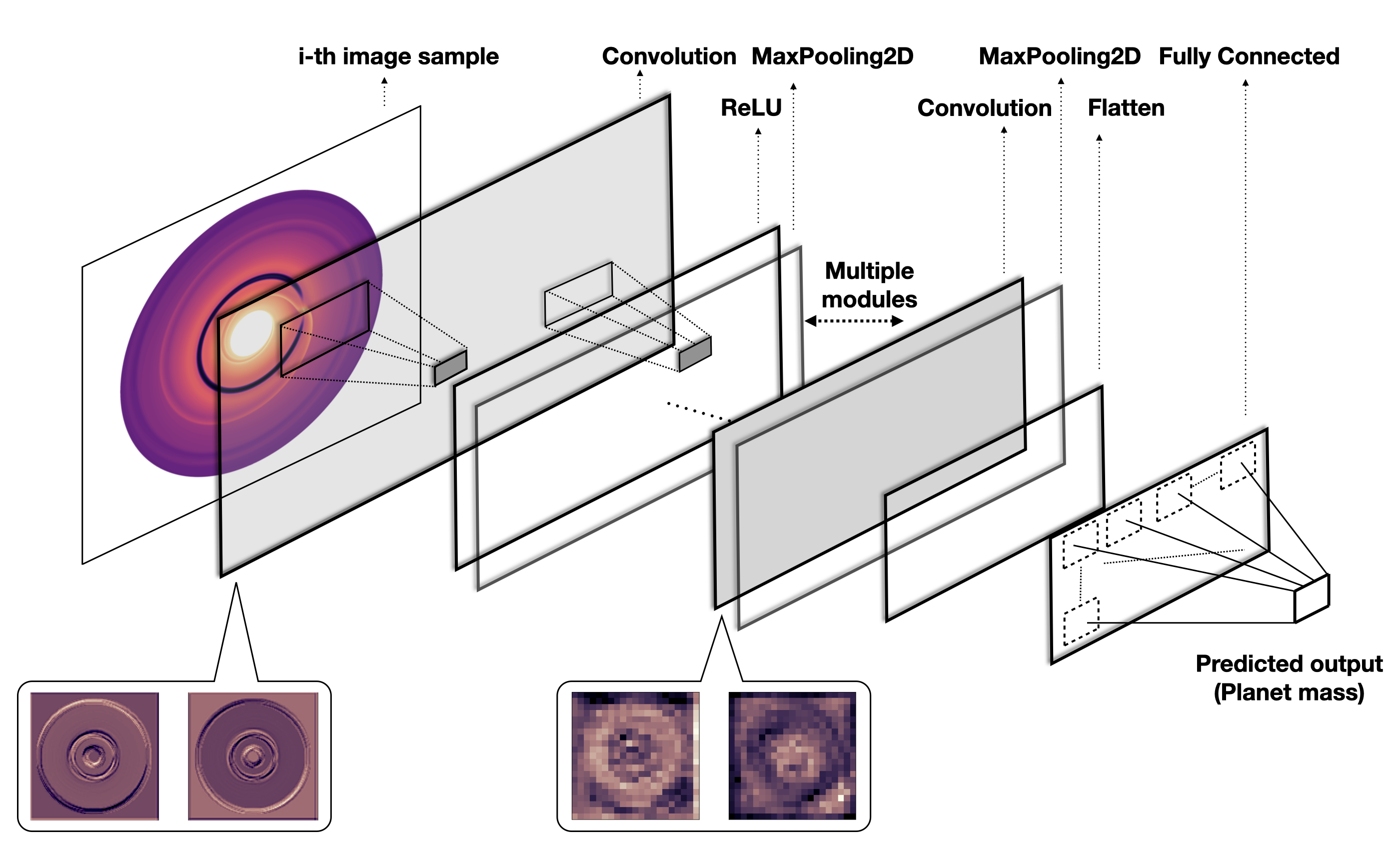}
\caption{A schematic diagram showing a typical/vanilla Convolutional Neural Network architecture along with the various hidden layers. It consists of multiple convolution and sub-sampling (MaxPooling2D) layers followed by a non-linearity (ReLU in this case). The characteristic feature of this architecture is the presence of a fully connected layer with linear activation for performing regression. The feature maps, showing both high and low level features, extracted by the convolutional filters are shown in the extended panels. The final output is the planet mass. Note, this schematic diagram is for illustration purpose only and the convolutional layers do not show the ResNet50 architecture used in DPNNet-2.0. While the input and the output layer remain the same, the convolutional layers can be swapped with different CNN architectures. For additional details see section \ref{architecture}}\label{fig:cnn}
\end{figure*}

\subsection{Simulation Setup and Parameter space}   
We use the \textsc{fargo3d} hydrodynamic code \citep{llambay19}, running in a 2D cylindrical geometry $(R,\phi)$, to simulate the dusty disk-planet system. The grid is uniformly spaced with $512 \times 512$ discretized cells. 
The computational domain spans from $R \in [0.4, 2.5]R_0$ and $\phi\in[0, 2\pi]$.  The radial boundaries are set to their initial equilibrium solution while we apply periodic boundary conditions in $\phi$. 
We damp waves generated by the planet to zero near the boundaries using the wave-killing module \citep{valborro06} to avoid reflections and interference with the disk bulk/morphology.
The planet is placed from the beginning of the simulation. We use a constant softening length of $r_s = 0.6h_0R_0$ for the planet's potential. The frame of reference rotates with the planet.

We consider the same parameter space, summarized in Table \ref{tab:input_data}, as used in Paper 1. This enable us to compare the relative performance between our previous ML model DPNNet and the current improved CNN model DPNNet-2.0. We follow each simulation for 3000 orbits which translates to about $\sim 1 \rm {Myr}$ at $45 \, \rm au$ around a $1 M_{\odot}$ star. The range of Stokes number considered here scale to particle radii ranging from $\sim 1 \rm mm $ to $ \sim 10 \, \rm cm $ assuming gas surface density $100 \, \rm g \, cm^{-2}$ and an internal grain density of $1 \,\rm g \, cm^{-3} $. We target super-Earth to Saturn-sized planets,  $M_{\rm P} \in [8, 120]M_{\oplus}$, as these are comparatively difficult to detect using conventional methods, thus making our deep learning model widely applicable.



\section{DPNNet-2.0} \label{sec-cnn}
In this section we briefly introduce  the CNN architecture and its implementation in designing the DPNNet-2.0 model. Further, we discuss the process of data acquisition, data pre-processing and DPNNet-2.0 training.   


\subsection{Convolutional neural network}
CNN is a type of feed-forward neural network consisting of convolutional layers followed by fully-connected layers \citep{rawat2017deep} that can be used for classification or regression problems. Figure \ref{fig:cnn} shows a typical CNN architecture along with the various layers within it. 
The crucial part of a CNN architecture is the introduction of convolutional layers, consisting of a set of neurons having shared weights. The convolutional layer can be viewed as a layer consisting of several neurons that analyses a small overlapping section of the input image/data. This idea was first introduced by  \cite{fukushima1982neocognitron} and  \cite{lecun1998gradient}, where the convolution operations were implemented in one or multiple layers in addition to the fully connected layers in the end.

CNNs have one major advantage over MLP based models. Due to the weight sharing as well as the small kernels, the system requirements for CNN is significantly reduced compared to MLP (with fully connected layers) performing general matrix multiplication.  This opened up the possibility of building faster, more cost effective, and deeper networks. They can perform exceptionally well in extracting features from an image (from low levels to higher levels) because of the hierarchical learning, enabled by stacking several convolutional filters having their own shared weights. These networks also comes with the advantage that they can automatically deal with spatial translations (rotation and scaling as well with some modifications).

Even though CNNs were initially designed to perform image classification \citep{krizhevsky2017imagenet} or object detection \citep{girshick2014rich}, in recent times these are also used for solving regression problems. This is implemented using a linear layer following the fully-connected layer in the end. Using this approach, commonly known as {\it vanilla deep regression}, state-of-the-art results are obtained in computer vision regression problems \citep{liu20163d,toshev2014deeppose,belagiannis2015robust}.
\\\\


\begin{figure*}
\centering
\includegraphics[ clip=True,scale=.2]{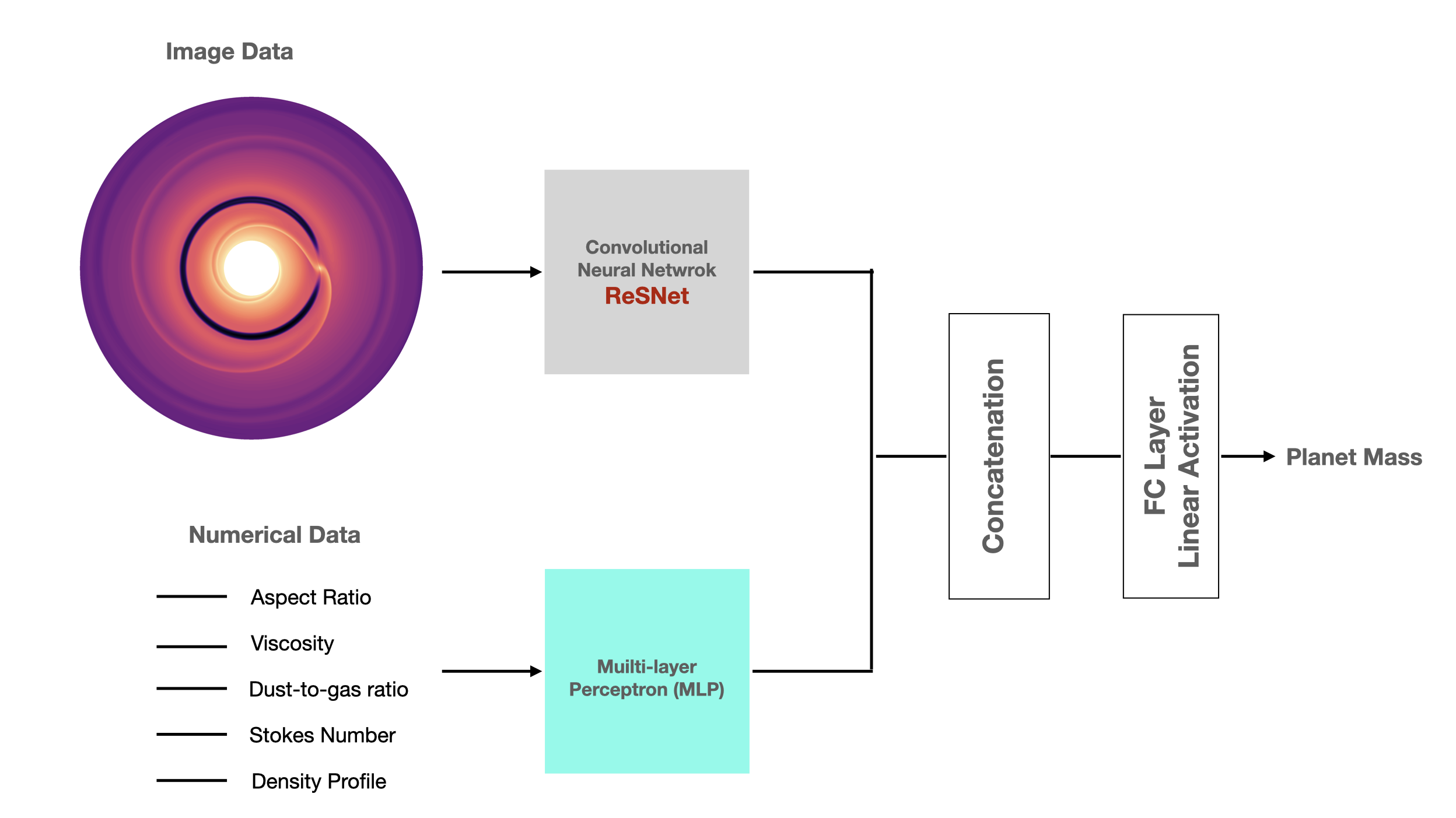} 
\caption{A schematic representation of the multi-input DPNNet-2.0 that can accept both the image and the disk characteristics as input. The image is fed to the CNN and the disk parameters propagate through the MLP model. The output from the CNN and the MLP is concatenated and passed through a fully connected (FC) layer and a linear activation layer. The final output is the planet mass. For more details refer to section \ref{DPNNet-2.0_hyb}}\label{fig:DPCNNet_hybird}
\end{figure*}

\subsection{DPNNet-2.0 architecture}\label{architecture}
DPNNet-2.0 uses a CNN architecture to predict planet mass using features extracted from input images directly. 
We tested the performance of DPNNet-2.0 with several well established 2D CNNs such as AlexNet \citep{krizhevsky2017imagenet}, VGG-16 \citep{simonyan2014very}, ResNet  \citep{he2016deep} as well as a {\it vanilla} CNN consisting of multiple convolution and pooling (sub-sampling) layers followed by a non-linearity (ReLU in this case). Our findings suggest that ResNet performs significantly better compared to the other CNN architectures and the training time is much shorter as well (see Figure (\ref{fig:bench_mark})). Based on this, the main analysis of this paper, as well as the benchmarking, is done using ResNet as the CNN block of the DPNNet-2.0 model.

The main reason for the superior performance of the ResNet architecture is the use of ``identity shortcut connections" that enables the construction of deeper networks while reducing the number of parameters. While AlexNet has 5 convolutional layers, VGG-16 has 19 layers and ResNet-50 has 50 layers. Simply increasing the number of layers to built a deeper network is infested with the problem of vanishing gradient, making them harder to train. For the case of ResNet, stacking several layers does not hinder the performance of the network because of the presence of the identity mappings (empty layers that does not do anything and just acts as a connection between two layers). Except the first convolutional layer of the ResNet architecture all other ones have a identity connection thus, making them {\it residual} convolutional blocks. These shortcuts or ``skip connections" allow the direct back-propagation of the gradient to earlier layers. 
For DPNNet-2.0 we use Adam optimizer, instead of RMSProp as used in \cite{aud20}, as it performs better incorporating the use of moving average of previous gradients in order to accelerate learning with inertia.


\begin{figure*}
\centering
\includegraphics[scale=.58]{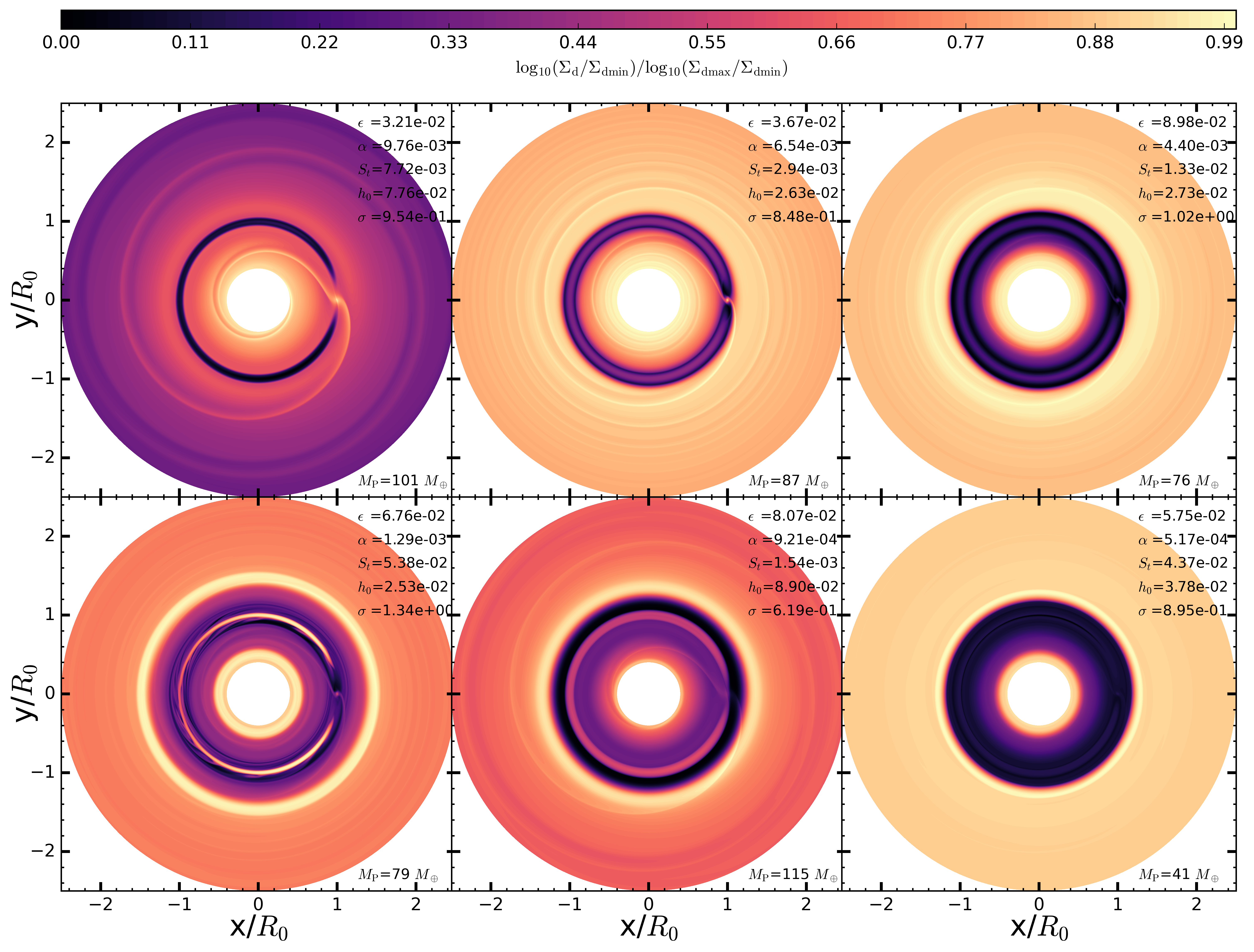}
\caption{The normalized dust surface density distribution after $3 \times 10^3$ orbits for different planet masses and different disk initial conditions indicated on the top right corner of each images. The top panel shows systems with a single gap with increasing width from left to right. The first two plots on the bottom panel have two gaps adjacent to a planet and finally the right most plot in the lower panel has a deep cavity. The mass of the embedded planet is indicated in each plot.  }\label{fig:disk_samples}
\end{figure*}

\subsection{Multi-input DPNNet-2.0}\label{DPNNet-2.0_hyb}
Multi-input DPNNet-2.0 is designed as an end-to-end model which can accept both images and disk features as input to predict the planet mass. This is crucial, as in addition to images each observed disk-planet system has its own physical characteristics, like disk profile, temperature, viscosity, as well particle size and metallicity. Thus, a complete model for planet-mass prediction needs to be sensitive not only to the disk morphology (observed from images) but also to the disk properties which are measured separately. 

This is achieved by developing an additional module that combines the CNN based architecture with the MLP network implemented in Paper 1. The MLP module is a fully connected feed forward neural network with two hidden layers, having 256 and 128 neurons, and an input layer with five feature variables. Figure \ref{fig:DPCNNet_hybird} gives the schematic of the network architecture of the multi-input DPNNet-2.0.

The image and the disk features are processed independently using the CNN and the MLP architecture in DPNNet-2.0 respectively. The concatenated output from the two branches serve as a input to a fully connected layers of neurons. Finally, it passes through a linear activation regression head to output the predicted planet mass. Note that we amend DPNNet (MLP model), such that it no longer requires gap width as input, as the gap features are characterized from the image directly by the CNN. Only the disk initial conditions ( $\alpha, h_{\rm 0}, \sigma, S_{\rm t}, \epsilon$) are considered. This allowed us to remove the uncertainties associated with DPNNet (user-defined gap width measurement) while retaining its flexibility.

\subsection{Data acquisition and pre-processing}\label{get_data}
We use Latin hypercube sampling (LHS) \citep{mck79,ima81} method to sample the parameter-space for initializing the disk-planet simulation. The LHS was implemented using (\textsc{pyDOE} package) to generate a uniform random distribution of the parameters. Each parameter value was centered within the sampling intervals. We run in total 1200 simulations initialized using the above parameter space.

For each simulation we generate dust surface density maps after the system has evolved substantially and have reached quasi-steady state. This corresponds to $\approx 2000$ orbits of evolution for most of our disk-planet simulation. Thus we generate dust density maps from 2400, 2600, 2800, and 3000 orbits of evolution. Thereby, for each simulation we have images from $4$ time instances resulting in a total of 4800 unique images for 1200 distinct simulations. This gives us a much larger sample size, which is necessary for better optimization of the CNN.



Figure \ref{fig:disk_samples} shows normalized dust surface density distribution for a set of sample simulations that are used to train the DPNNet-2.0. The most distinct and visible features are the presence of dust gaps and rings. These annular structure can be broadly categorized as a) single deep gap of increasing width (left to right) as seen in the top panel, (b) two gaps adjacent to a planet (first two plots from left in the lower panel) (c) a deep cavity as evident on right most plot in the lower panel in Figure \ref{fig:disk_samples}. For details about the characteristic about the gaps refer to section (3.2) in Paper 1.


The images are initially filtered to eliminate runs that have undetectable gaps or more than two gaps. After the initial screening we are left with 2945 images combined from all the four different time instances. We crop the images to remove the axis and the color bars. Then they are resized to exactly match the required resolution, which is this case was $512 \times 512$. The cropped images are further pre-processed before feeding them as input to the network. The pixel values of the images ranges between 0 and 255. We normalize each pixel values within the range $0-1$, as inputs with larger values can slow down/disrupt the training process. 
Furthermore, we implement a standardization technique where the pixel values are distributed normally with a mean of 0 and standard deviation of 1.

Each image is associated with a set of input parameters $M_\mathrm{P}, \alpha, h_{\rm 0}, \sigma, S_{\rm t}, \epsilon$. We assign each image a target variable planet mass $M_{\rm P}$. For the single input DPNNet-2.0 the images serves as the sole input amd the planet mass is the targeted output. However, for the multi-input DPNNet-2.0, we assign each simulation a set of feature variables ($\alpha, h_{\rm 0}, \sigma, S_{\rm t}, \epsilon$) and the corresponding image as input. The target variable is the planet mass $M_{\rm P}$. The data for each feature variable is normalized using standard (z-score) scaling by subtracting the mean and scaling it by the standard deviation. 



\begin{figure*}[ht] 
\centering
\includegraphics[ clip=True,scale=.9]{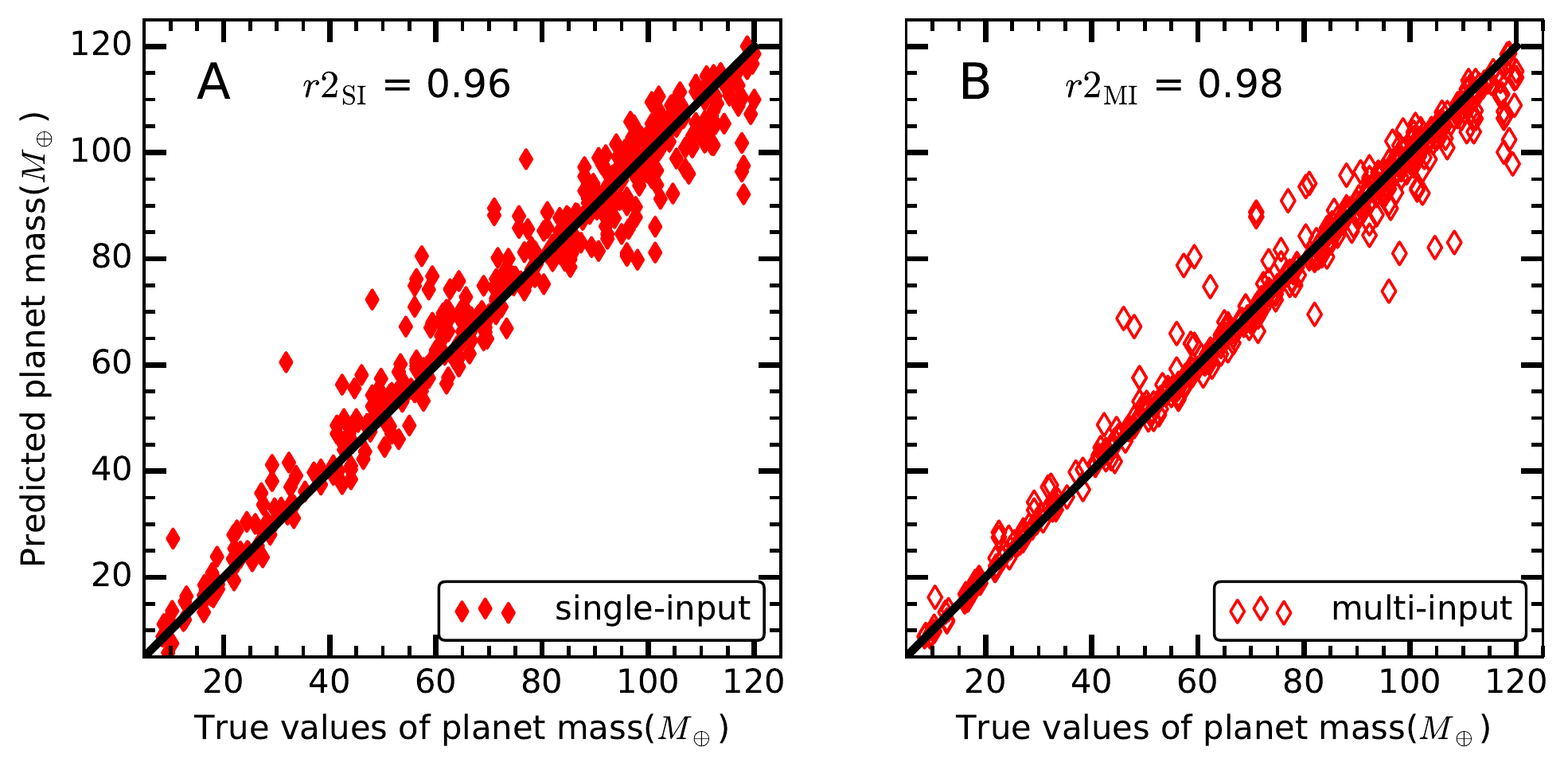}
\caption{Correlation between the simulated planet mass and the predicted planet mass (in the units of $M_{\Earth}$) obtained from the single-input (only images) and multi-input (images and disk parameters) DPNNet-2.0 model. The $r2-$ scores indicate the goodness of fit. }\label{fig:RESNET_pre}
\end{figure*}

\subsection{DPNNet-2.0 Training}
The CNN and the MLP architectures are implemented using the deep learning API, Keras, which acts as an interface for TensorFLow \citep[][]{tensorflow2015-whitepaper}. 
For the purpose of training and testing the model we split the dataset into two blocks, a training set consisting of 85\% and a testing set consisting of 15 \% of the simulation data. The training set is further split to keep 15\% of the data aside for model validation. There are two separate training processes for the two modules of the DPNNet-2.0. For the single input CNN based module only the pre-processed simulated images from the training set are fed into the network as input. 
The multi-input module accepts both the pre-processed simulated images and the corresponding normalized feature variables  ($\alpha, h_{\rm 0}, \sigma, S_{\rm t}, \epsilon$). In both the cases the target variable, planet mass $M_{\rm P}$, is the same. The resolution of the input images are fixed to $512 \times 512$.

The training process works iteratively where the weights are adjusted after each step until they are optimized. For each iteration a subset of the training dataset is fed into the network (batch size = 20) to minimize the computational cost during the optimization process. We use an adaptive learning rate $10^{-5}$ with a small decay of $1/2000$. The single and multi-input modules are trained independently for several epochs while monitoring the validation loss, mean square error (MSE), after each iteration. We implement early stopping and use the callback function to stop the training when the model is stable and reaches optimal performance (minimum MSE).
This is done to avoid over-fitting or under-fitting of the model. A small number of epoch would result in an under-fit model whereas a large number of epoch might lead to over-fitting.
Figures \ref{fig:CNN_MSE} shows the behavior of the loss function (in terms of MSE) as a function of epochs during the training process till it stabilizes for both the modules.
Note, the training process is almost independent of the batch size due to the use of ResNet architecture, which is shown to be insensitive to the specific choice of batch size \citep{dblp}.
The training is accomplished with GPU machines \textsc{RTX 8000} and \textsc{Tesla P100} available in the super-computing clusters at ASIAA. For the initial code development and testing we used M1-ARM based chip of Apple and the non-paid version of Google Colab. However, both Google Colab and M1 had limited GPU memory and could only be used for low resolution image. Figure \ref{fig:bench_mark}, gives the benchmarking of the performance for the different hardware against the use of various CNN architecture.


 


\section{Results} \label{sec-results}
Once the training process is complete, we have two trained modules of DPNNet-2.0 that can be deployed to predict planet masses from disk images. The single-input module accepts simulated images as input, while the multi-input one additionally considers the feature variables. We deploy both the modules on the test dataset to quantify the network's performance when applied to unseen data. We compare the network's accuracy in terms of the r2-score  as well as the root mean square error (RMSE) for the single and the multi-input modules.





The test dataset consists of randomly chosen 429 sample simulated images along with the corresponding feature variables and the true values of the planet mass  (as used in the underlying hydrodynamic simulation). These images are fed into the single-input DPNNet-2.0 to predict the planet mass. Figure \ref{fig:RESNET_pre}-A illustrates the correlation between the predicted planet mass (in Earth-mass $M_{\oplus}$ units) and the actual values of the planet mass used in the simulations. The predicted planet masses mostly lie along the black line indicating a strong correlation with a r2-score of $r2_{\rm SI} = 0.96$. We further estimate the RMSE and the mean absolute error (MAE) to quantify the model's performance. DPNNet-2.0 with single-input has a RMSE of 6.2 $M_{\Earth}$ and a MAE of $4.3 M_{\Earth}$ when applied to the test dataset. 

On implementing the multi-input module we have an improved model performance. Figure \ref{fig:RESNET_pre}-B shows the correlation between the predicted and the simulated planet mass. The correlation is much tighter with $r2_{\rm MI} = 0.98$ which is an improvement compared to the single-input module. The RMSE and the MAE are $4.6 M_{\Earth}$ and $2.5 M_{\Earth}$ respectively which are comparatively lower.

To further reduce statistical fluctuations due to data sampling, we apply $k$-Fold cross validation, which is a re-sampling process, using the Python \textsc{scikit-learn} library. The data is shuffled randomly and divided in $k=7$ folds or groups of equal size. A single fold is used for testing while remaining $k-1$ folds are used to train the model. The process is repeated $k$ number of times, each time selecting a different group to test and the remaining groups to train. The mean of the RMSE values obtained from each iteration is used as a measure of the model performance. Figure (\ref{fig:RMSE}) shows the variation of the mean RMSE (left panel) and the mean r2-score (right panel) as function of image resolution for both the single and multi-input DPNNet-2.0. At the maximum image resolution of $512 \times 512 $ we measure a mean RMSE of ($7.5M_{\Earth}, 4.9M_{\Earth}$)  and mean r2-score of (0.94, 0.97) for the single and multi-input models respectively. These values can be considered as the prediction uncertainty of DPNNet-2.0. Thus the use of image  based  characterization enables DPNNet-2.0 to perform significantly better compared to the MLP based model (DPNNet) in Paper 1, which reported a RMSE of $12.5M_{\Earth}$ and MAE of $7.9M_{\Earth}$ when applied to the test dataset.

\begin{figure}[ht]
\centering
\includegraphics[width=3.5in,trim=03mm 00mm 00mm 00mm, clip=True]{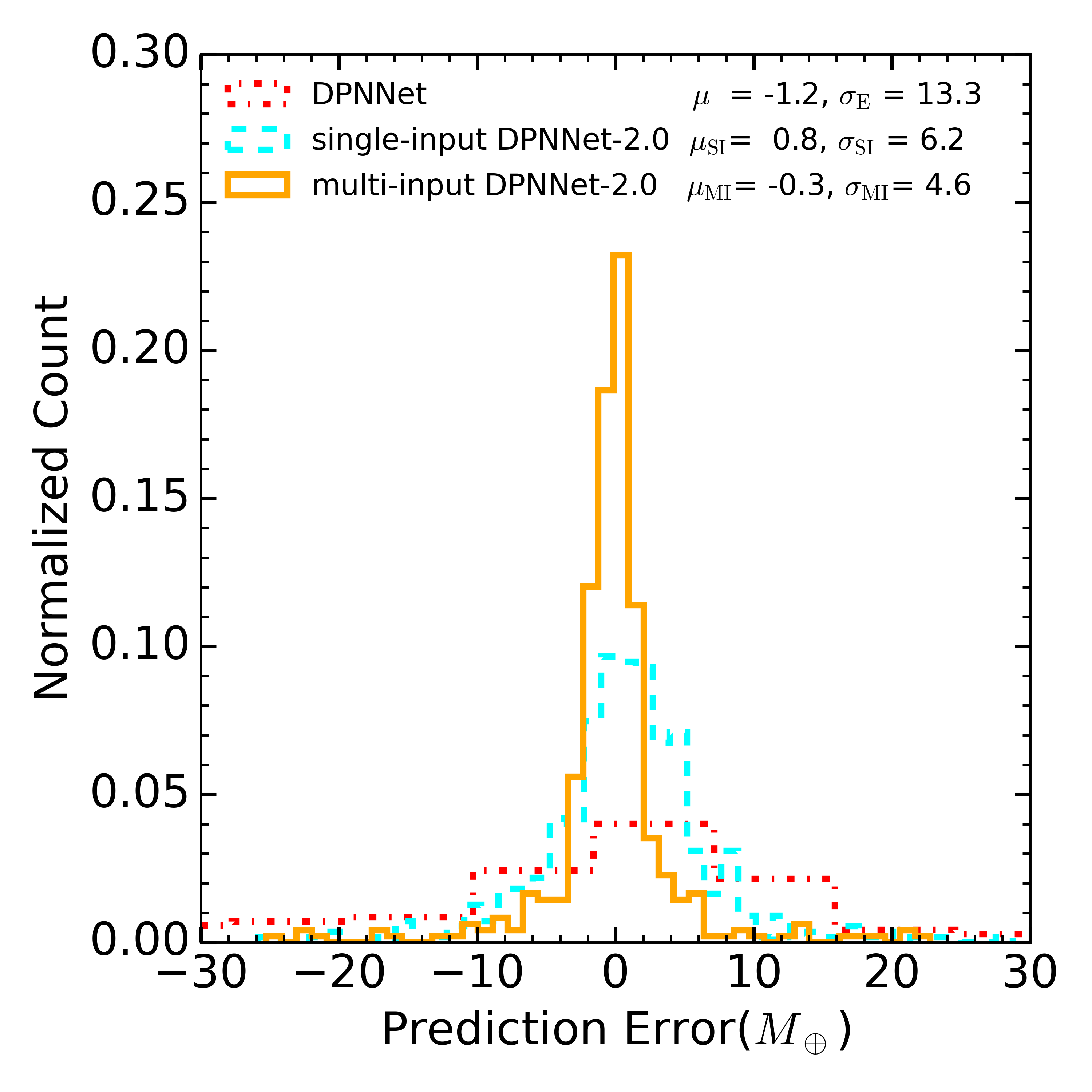}
\caption{Distribution of the prediction error $M_{\rm P,simulated}- M_{\rm P,predicted}$ for DPNNet, single-input  and multi-input DPNNet-2.0.}\label{fig:model_error}
\end{figure}

This is further illustrated in Figure (\ref{fig:model_error}), where we plot the prediction error $M_{\rm P,simulated}- M_{\rm P,predicted}$ for each of the models. The error follows a normal distribution. The mean for DPNNet-2.0 ($|\mu|_{\rm SI} = 0.8,|\mu|_{\rm MI} = 0.3$) is more centered towards zero and the standard deviation ($\sigma_{\rm SI} = 6.2,\sigma_{\rm MI} = 4.6$) is also lower compared to DPNNet ($|\mu| = 1.2,\sigma = 13.3$) indicating a tighter constrain for the predicted planet mass. Thus image based characterization of disk images using CNN significantly brings down the prediction uncertainly/error. DPNNet-2.0 outperforms its predecessor  DPNNet in term of both predictive capability as well as the flexibility to accept images as direct input.


\section{Discussion and conclusion} \label{sec-discussion}
In this paper we introduce part I of the CNN-based DPNNet 2.0 architecture to predict the mass of unseen exoplanets from disk images. Currently this model is trained and tested on images generated from hydrodynamic simulations of disk-planet interaction. We demonstrate the high accuracy and the efficiency of the  network to predict the planet mass from the simulated images. This provides the required framework for extracting the mass of exoplanets from observed images, which is a work in progress, where we need to train the model with synthetic images (DPNNet 2.0 part II, Auddy et al.). In part I we primarily focus on building the CNN based architecture and demonstrate its functionality based on simulated images.

In \cite{aud20}  we developed DPNNet, a MLP based model that accepts a set of disk parameters and predicts the planet mass. This required measuring the gap width explicitly from each of the observed images, which is often prone to additional uncertainties (see section (\ref{DPNNet})). In this work we exploit the power of computer vision, implementing a CNN architecture, to learn in a more direct way the underlying relationship between the features observed in the simulated images of PPDs hosting an embedded planet. A CNN architecture can extract features from a given image using the various convolutional filters present within the network.
This makes it possible to map the connection between the various disk structures, most prominent being the dust gap, and the complex disk-planet interaction.



Equipped with a CNN at its core, DPNNet-2.0 predicts the mass of the unseen planet from the disk morphology. It extracts the features directly from images, thereby minimizing the errors associated with different user-defined measurement probes. Compared to the results of Paper 1 having a mean RMSE of $\pm 12.5 M_{\Earth}$, the single-input DPNNet-2.0 yields a much tighter constrain to the predicted planet mass with a reduced mean RMSE of $\pm 7.4 M_{\Earth}$.
This can also be attributed to the fact that in addition to considering the visible gap width, DPNNet-2.0 takes in account other morphological characteristics (like gap depth, asymmetries like vortices and spirals) which are imprint of disk-planet interaction as observed in simulated disk images. This makes DPNNet-2.0 much more effective in constraining the properties of unseen planets, like planet mass.

The multi-input module in DPNNet-2.0 is designed to accept disk features in addition to dust surface density simulated images. It inherits the MLP module from DPNNet (Paper 1), which  gives it the flexibility to accept disk features, like $\alpha, h_{\rm 0}, \sigma, S_{\rm t}, \epsilon$ along with images. This allows the network to further reduce the uncertainty in the predicted planet mass. It yields a mean RMSE of $\pm 4.9 M_{\Earth}$, which is an improvement over the single-input model. Thereby, multi-input DPNNet-2.0 retains the beneficial part of DPNNet while improving its capacity to better extract features from the simulated images.

\begin{figure}[ht]
\centering
\includegraphics[scale=.65,]{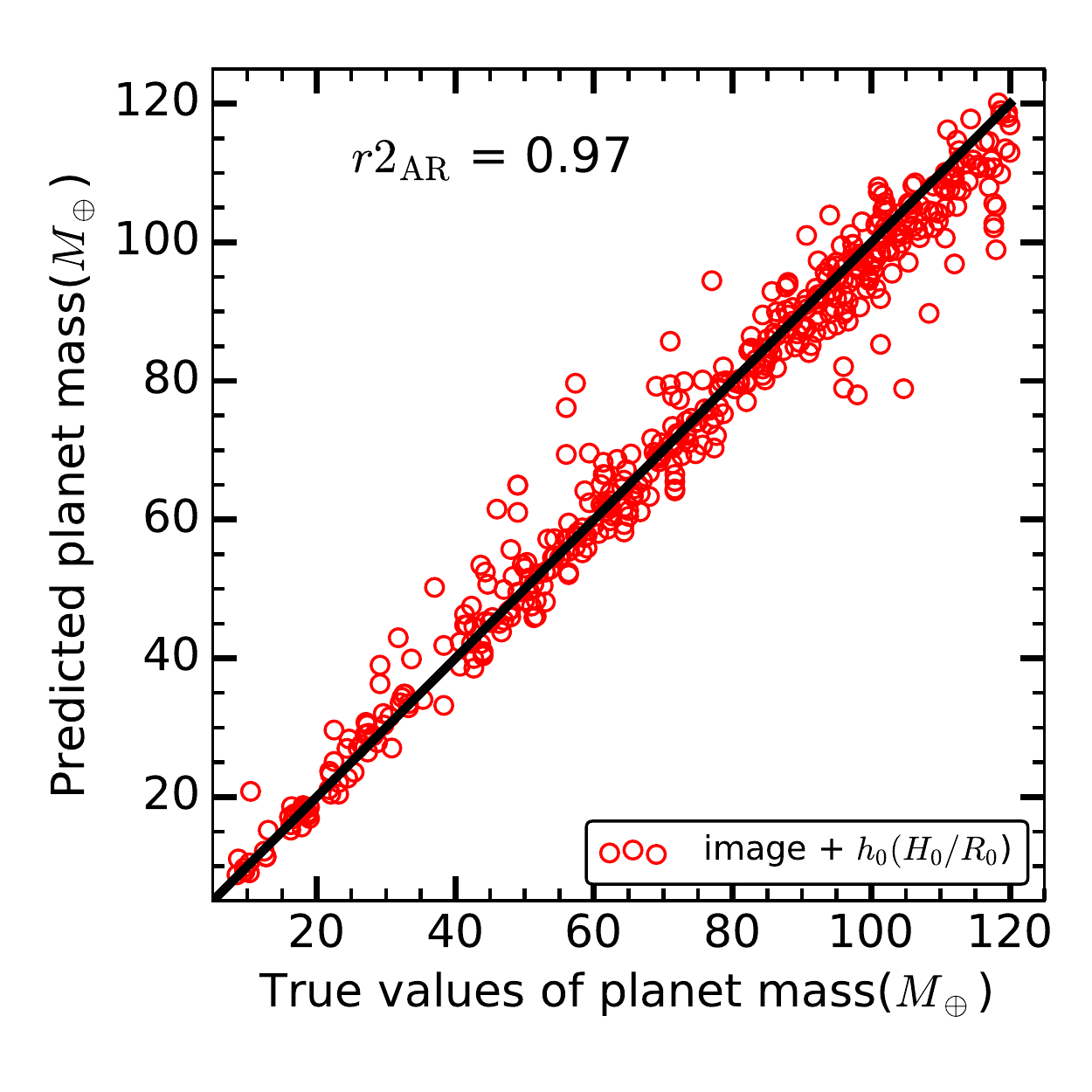}
\caption{Correlation between the actual and the predicted mass when using simulated image as well as the disk aspect ratio $h_{0}$ as inputs.}\label{fig:AR_predicted}
\end{figure}

However, many of the the disk parameters that are required for the multi-input DPNNet-2.0 are often not well constrained from observation. One usually considers canonical values when precise measurement is not possible. For example, it is common to consider a disk to be  weakly turbulent, $\alpha \approx 10^{-3}$, with  particle abundance (dust-to-gas ratio) of $\sim 1\%$, comprising of millimeter sized dust grains. 
 On the other hand, disk temperatures can be constrained with more confidence (Y.-W. Tang, private communication), which
is used to estimate the disk aspect ratio $h_{0}$. Thus, as a test we train an addition multi-input model where we drop all the feature variables except the disk aspect ratio $h_{0}$. Once trained we test the network on the same test data, but this time the network only accepts simulated disk image and the corresponding $h_{0}$. Figure (\ref{fig:AR_predicted}) gives the correlation between the predicted planet mass and the the simulated values. The r2-score is $r2_{\rm  SI}<r2_{\rm AR}<r2_{\rm MI}$ implying an increase in accuracy compared to single-input but  a decrease compared to multi-input models. This is indicative of the fact that with addition of more disk parameters as input, along with simulated image, the prediction uncertainty gets reduced.


In the next sections we will highlight the limitations associated with limited training data, CNN architecture and discuss scopes of improvement in future.




\section{Network Limitations and Future scopes}
The key to an effective well-trained model is a generous dataset that includes a broad parameter space and captures a wide diversity in disk morphology induced by disk-planet interaction. A trained network is as good as its training dataset, which ideally must encompass all possible scenarios. However, it is not only challenging to design simulations which does that, it is equally expensive to generate the data. In this paper we only use simulated images and explore a narrow parameter space due to finite computational resources. 

The ultimate objective of DPNNet-2.0 is to predict planet mass from observed images. However, due to limited availability of observed PPD images the model cannot be solely trained with observed data.
Thus DPNNet-2.0 is trained and tested on simulated images to verify the capability of the framework. Part I of the DPNNet-2.0 model, this paper, mainly focuses on the model build and the network architecture. Though at its current state, this model has its limitations, this novel approach opens an unique direction of using computer vision for parameter estimation (for e.g, planet mass) from observed disk images. Part II of this work will focus on developing more realistic training datasets (e.g., synthetic images, wider parameter space) to improve the robustness and applicability of the model on observed images.

 In DPNNet-2.0 part II (Auddy et al., in preparation) we use synthetic observations as the training dataset. This requires 3D disk models obtained by running 3D hydrodynamic simulations or by “puffing up” 2D disk models. In either case, we follow up with radiative transfer calculations using a variety of dust opacities and protostellar properties, and put them through a synthetic image generator to mimic interferometric observations \citep[][]{dong15}. Once the dataset is generated we will simply re-train the DPNNet-2.0 model with these synthetic observations without changing the overall architecture of the network. 

For deploying DPNNet-2.0 to observations, one should also  account for additional factors such as the viewing angle, dust opacity, observing wavelength, multiple gaps, stellar properties, to name a few. These dependencies will generally increase the training data set. However, some of these issues can be addressed without any change to the DPNNet-2.0 architecture. For example, although DPNNet-2.0 is trained on face-on images, in principle it can be applied to observations at other viewing angles by deprojecting it. It is, in fact a standard procedure to deproject observed ALMA disk images \citep[for example see][]{Hua18a,Hua18b,pinte20} with varying viewing angles (i.e. inclined and/or rotated with respect to the observer) to obtain face-on images. One can directly feed these deprojected images to DPNNet-2.0 as input without altering the training process. In DPNNet-2.0 part II and future follow-ups we will address these issues in more detail, overcoming the limitations of computational resources, and work towards improving the efficiency of the model.

\section{Code Availability }
The codes used for developing DPNNet-2.0 are available at the GitHub software repository, at this link:

https://github.com/sauddy/DPNNet-2.0

The codes for the DPNNet in \citep{aud20} are also available at 

https://github.com/sauddy/DPNNet

\section*{Acknowledgments}
We thank the anonymous referee for the constructive comments and feedback.
MKL is supported by the Ministry of Science and Technology of Taiwan (grants 107-2112-M-001-043-MY3, 110-2112-M-001-034-, 110-2124-M-002-012-) and an Academia Sinica Career Development Award (AS-CDA-110-M06). Numerical simulations
were performed on the TIARA cluster at ASIAA, as well
as the TWCC cluster at the National Center for High-performance Computing (NCHC). This study was supported in part by resources and technical expertise from the Georgia Advanced Computing Resource Center, a partnership between the University of Georgia’s Office of the Vice President for Research and Office of the Vice President for Information Technology.

\begin{figure}[ht]
\centering
\includegraphics[scale=.50]{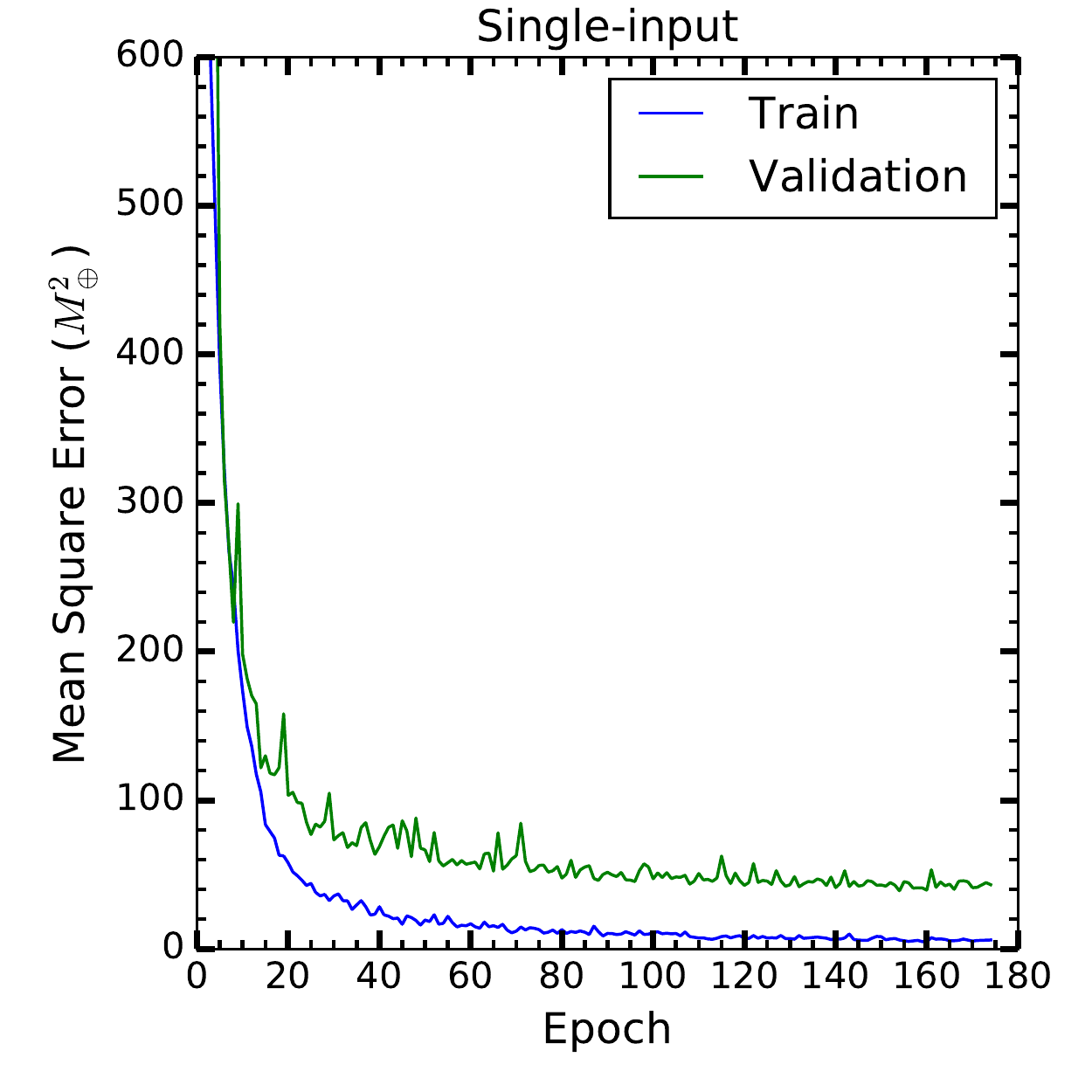}
\includegraphics[scale=.50]{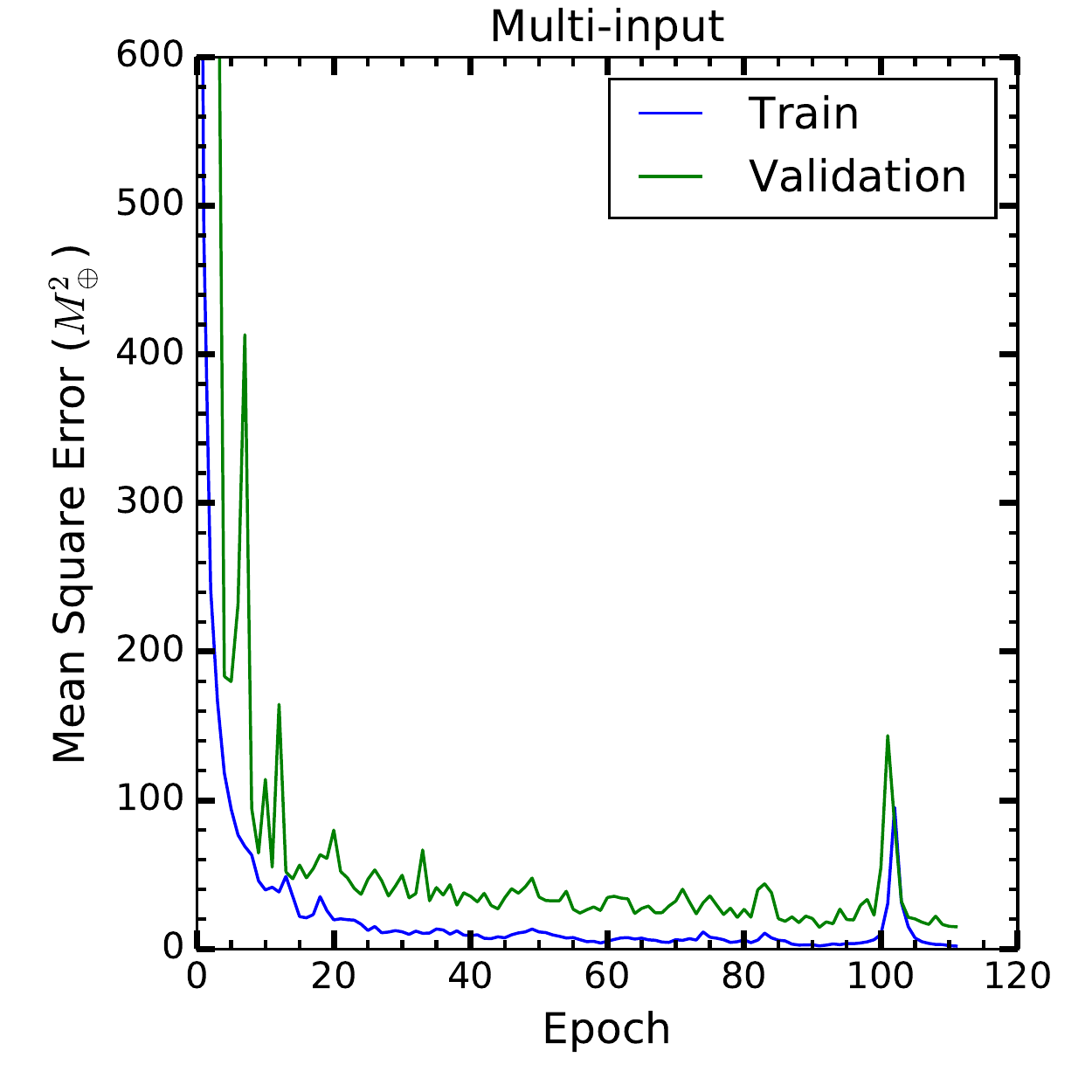}
\caption{The training and validation loss is shown as a function of training epoch for the single as well as the multi-input DPNNet-2.0 model  }\label{fig:CNN_MSE}
\end{figure}
\begin{figure}
\centering
\includegraphics[width=5.0in,trim=00mm 00mm 00mm 00mm, clip=True]{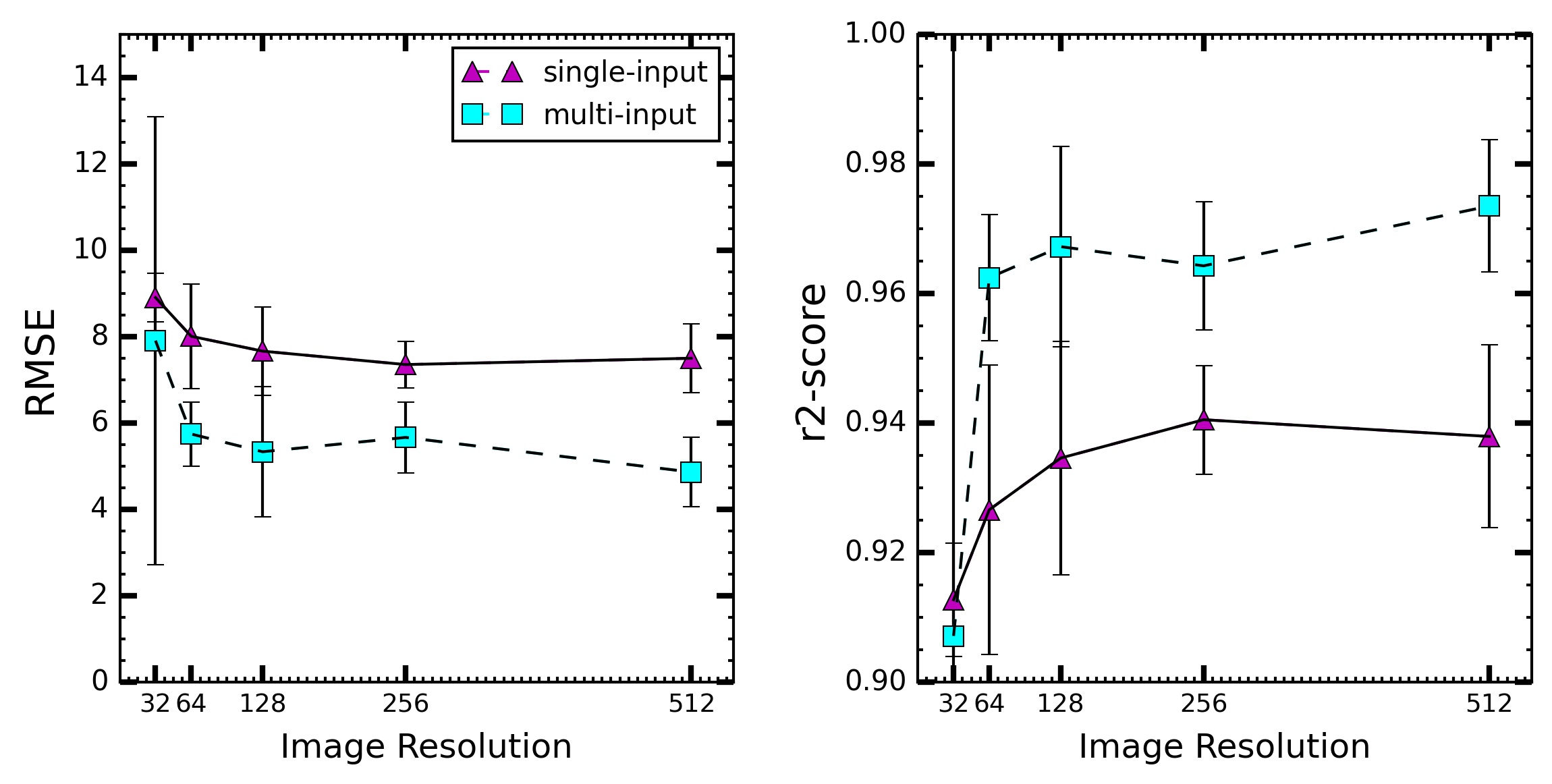}
\caption{Left: The variation of the mean value of RMSE with the increasing resolution of images obtained from $k-$Fold cross validation. Right: The corresponding mean value of r2-score indicating the goodness of fit between the predicted and the simulated planet mass. The error bar indicated the standard deviation incurred due to sampling. }\label{fig:RMSE}
\end{figure}
\begin{figure}
\centering
\includegraphics[scale=.4]{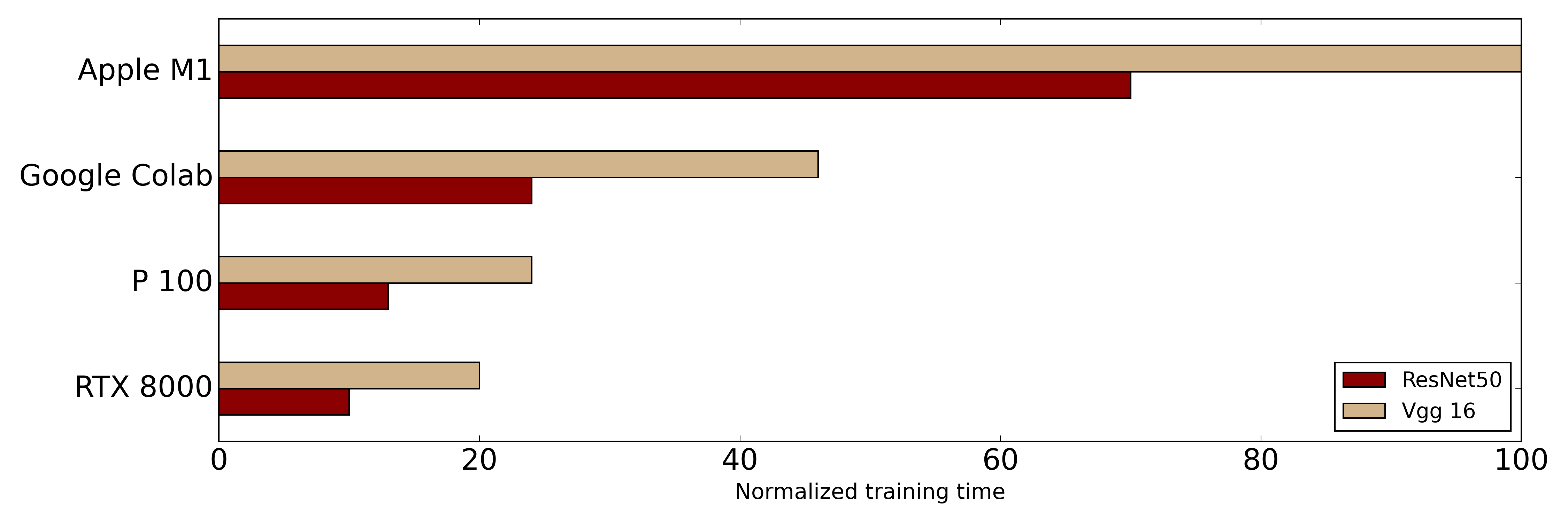}
\caption{Comparison of different GPU hardware in terms of training time for the ResNet 50/Vgg-16 architecture used in the DPNNet-2.0 model}\label{fig:bench_mark}
\end{figure}
\begin{figure}[ht] 
\centering
\includegraphics[scale=.4]{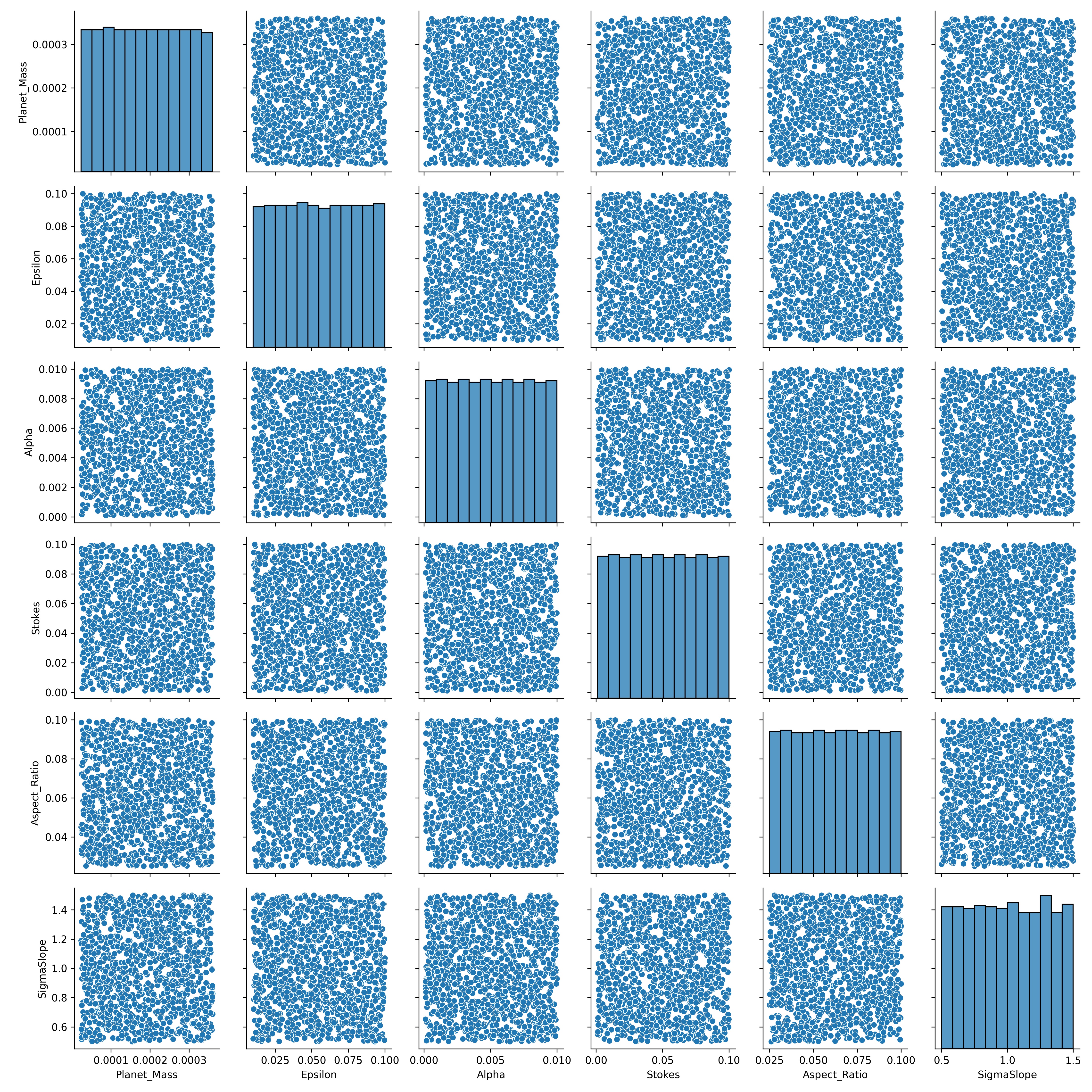}
\caption{Pairplot representing the distribution of the input parameters (initial condition) sampled using LHS.The correlation between each quantities is demonstrated in each panel. The histograms of the different quantities are also represented in the diagonal. The LHS was implemented using (pyDOE package) to generate a uniform random distribution of the parameters. Each parameter value was centered within the sampling intervals}\label{fig:LHS_params}
\end{figure}

{
\appendix

Figure (\ref{fig:CNN_MSE}) shows the training and the validation loss (MSE) as a function of training epoch for the single and the multi-input models respectively. In Figure (\ref{fig:RMSE}) we show the variation of the mean RMSE and the r2 score for different image resolutions. 
Figure (\ref{fig:bench_mark}) gives the relative computation cost for both ReSNet50 and Vgg-16 using different GPU hardware. Figure (\ref{fig:LHS_params}) is a pairplot representing the distribution of the input parameters sampled using LHS.
}

\newpage
\bibliography{planet_cnn}{}
\bibliographystyle{aasjournal}

\end{document}